\documentclass[preprint,aps,showpacs,nofootinbib,preprintnumbers,amsmath,amssymb]{revtex4-1}
\usepackage{}
\usepackage{epsfig}
\usepackage{subfigure}
\usepackage{dcolumn}
\usepackage{bm}
\usepackage[usenames ,dvipsnames]{xcolor}
\usepackage{slashed}
\usepackage{graphicx,color}
\usepackage{centernot}
 \begin{document}
\title{Determination of $|V_{ub}|$ from exclusive baryonic $B$ decays}

\author{Y.K.~Hsiao and C.Q.~Geng}
\affiliation{
Chongqing University of Posts \& Telecommunications, Chongqing, 400065, China\\
Physics Division, National Center for Theoretical Sciences, Hsinchu, Taiwan 300\\
Department of Physics, National Tsing Hua University, Hsinchu, Taiwan 300
}
\date{\today}

\begin{abstract}
We use the exclusive baryonic $B$ decays to determine the 
Cabibbo-Kobayashi-Maskawa (CKM) matrix element $V_{ub}$.
From  the relation $|V_{ub}|^2/|V_{cb}|^2=({\cal B}_\pi/{\cal B}_D){\cal R}_{ff}$
based on $B^-\to p\bar p \pi^-$ and $\bar B^0\to p\bar p D^0$ decays,
where $|V_{cb}|$ and ${\cal B}_\pi/{\cal B}_D\equiv{\cal B}(B^-\to p\bar p \pi^-)/{\cal B}(\bar B^0\to p\bar p D^0)$
are the data input parameters, while ${\cal R}_{ff}$ is the one fixed by the $B\to p\bar p$ transition  matrix elements,
we find $|V_{ub}|=(3.48^{+0.87}_{-0.63}\pm 0.40\pm 0.07)\times 10^{-3}$ with the errors corresponding
to the uncertainties from ${\cal R}_{ff}$, ${\cal B}_\pi/{\cal B}_D$ and $|V_{cb}|$, respectively.
Being independent of the previous results,
our determination of $|V_{ub}|$ has the central value close to those from 
the exclusive $\bar B\to \pi\ell\bar \nu_\ell$ and $\Lambda_b\to p\mu^-\bar \nu_\mu$ decays, 
but overlaps the one from the inclusive $\bar B\to X_u \ell\bar \nu_\ell$ with the current uncertainties.
The extraction of $|V_{ub}|$ in the baryonic $B$ decays is clearly very useful for the complete 
determination of the CKM matrix elements as well as the exploration of new physics.
\end{abstract}

\maketitle
\section{introduction}
In the standard model (SM), the unique physical phase in the 
$3\times 3$ unitary Cabibbo-Kobayashi-Maskawa (CKM) matrix~\cite{CKM1,CKM2}
provides the only source  for  CP violation.
However, it is known that this CP phase is not sufficient to 
solve the mystery of the matter-antimatter asymmetry in the universe.
To test the SM and look for other CP violation mechanisms, 
many CP violating processes,
proceeding through  the CKM matrix element
$V_{ub}=|V_{ub}|e^{-i\gamma}$ with $\gamma$ the CP phase,
have been extensively explored 
by both experimental and theoretical studies.
Nonetheless, with $\gamma$ 
more precisely analyzed from the present data~\cite{pdg}, 
the $|V_{ub}|$ determination is not conclusive.
In particular,
the experiments in the inclusive $\bar B\to X_u\ell\bar \nu_\ell$
and  exclusive $\bar B\to \pi \ell\bar \nu_\ell$ decays give~\cite{pdg}
\begin{eqnarray}
|V_{ub}|&=&(4.41\pm 0.15^{+0.15}_{-0.17})\times 10^{-3}\,,
\label{Va}\\
|V_{ub}|&=&(3.28\pm 0.29)\times 10^{-3}\,,
\label{Vb}
\end{eqnarray}
respectively,  where the first result has 3$\sigma$ deviation from the second one.
This is the well-known long-standing tension between $V_{ub}$ measured by inclusive and exclusive 
decays at the $B$-factories, 
which triggers the theoretical studies in the SM~\cite{Kang:2013jaa,Feldmann:2015xsa}.

To resolve the problem,
it has been proposed that there exists some new physics, such as
the right-handed quark current with the form of $\bar u\gamma_\mu (1+\gamma_5) b$
in $\bar B\to X_u\ell\bar \nu_\ell$~\cite{Crivellin:2009sd,Buras:2010pz},
but not supported by the test of $B\to \rho \ell\bar \nu_\ell$~\cite{Crivellin:2014zpa}.
It is also not sustained by the recent measurement of 
$|V_{ub}|=(3.27\pm 0.15\pm 0.17\pm 0.06)\times 10^{-3}$~\cite{LHCb_Vub}
in the exclusive baryonic decay of $\Lambda_b\to p \mu^-\bar \nu_\mu$,  which
contains  both contributions from 
the vector and axial-vector quark currents as $\bar B\to X_u\ell\bar \nu_\ell$.
Clearly, the resolution of the dual nature of 
$|V_{ub}|$ in Eqs.~(\ref{Va}) and  (\ref{Vb}) is one of the most important tasks in particle physics
and it would lead to physics beyond the SM.

The exclusive baryonic $B$ decays is worthwhile to have its own version for the extraction of $|V_{ub}|$, 
which can be independent of the previous ones from  
$\bar B\to \pi \ell\bar \nu_\ell$ and $\Lambda_b\to p \mu^-\bar \nu_\mu$.
For $\Lambda_b\to p \mu^-\bar \nu_\mu$, the extraction of $|V_{ub}|$ relies on
 the relation of $|V_{ub}|^2/|V_{cb}|^2=[{\cal B}(\Lambda_b\to p\mu\bar \nu_\mu)/
{\cal B}(\Lambda_b\to \Lambda_c^+\mu\bar \nu_\mu)]{\cal R}_{FF}$~\cite{LHCb_Vub}
with ${\cal R}_{FF}$ as the ratio of 
the $\Lambda_b\to p$ and $\Lambda_b\to\Lambda_c$ transition
form factors calculated in the lattice QCD~\cite{Detmold:2015aaa}.
Likewise, by connecting
$B^-\to p\bar p \pi^-$ and $\bar B^0\to p\bar p D^0$ decays
that proceed through $b\to u\bar u d$ and $b\to c\bar u d$ at the quark level, respectively,
we obtain $|V_{ub}|^2/|V_{cb}|^2=({\cal B}_\pi/{\cal B}_D){\cal R}_{ff}$
with ${\cal B}_\pi/{\cal B}_D\equiv{\cal B}(B^-\to p\bar p \pi^-)/{\cal B}(\bar B^0\to p\bar p D^0)$
and ${\cal R}_{ff}$  the parameter related to the hadronic effects including 
those from the $B\to p\bar p$ transition matrix elements.
Note that  the momentum dependences of these transition elements
have been well studied in the literature~\cite{HouSoni,Chua:2002wn,Geng:2006wz,Geng:2006jt}
to explain the threshold effect in the baryonic B decays with  fully accounted theoretical uncertainties.
On the other hand, the decay of $\bar B\to \pi \ell\bar \nu_\ell$
can not be isolated from the uncertainty caused by
the momentum dependences of the form factors in the $\bar B\to \pi$ transition,
calculated in different QCD models~\cite{Vub_Belle,Vub_Babar}.
Besides, since $|V_{ub}|^2/|V_{cb}|^2=({\cal B}_\pi/{\cal B}_D){\cal R}_{ff}$
receives the contributions from both vector and axial vector currents, 
it can also be used to test new physics in the form of the axial vector current.
It is clear that once ${\cal R}_{ff}$ is obtained in the baryonic $B$ decays, one can determine 
$|V_{ub}|$ from $|V_{ub}|^2/|V_{cb}|^2=({\cal B}_\pi/{\cal B}_D){\cal R}_{ff}$, 
which is independent of the previous cases.

\section{data analysis}
\subsection{Amplitudes}
In terms of the quark level effective Hamiltonian for the 
$b\to c \bar u d$ and $b\to u \bar u d$ transitions,
the amplitudes of the $B^-\to p\bar p \pi^-$ and  
$\bar B^0\to p\bar p D^0$ decays can be derived 
as~\cite{Geng:2006wz,Geng:2006jt,Chen:2008sw,Hsiao:2013dta}
\begin{eqnarray}\label{amp1}
{\cal A}(B^-\to p\bar p \pi^-)&\simeq &i\frac{G_F }{\sqrt 2}V_{ub}V_{ud}
a_1 f_\pi \langle p\bar p|\bar u\centernot p (1-\gamma_5)b|B^-\rangle\,,\nonumber\\
{\cal A}(\bar B^0\to p\bar p D^0)&=&i\frac{G_F }{\sqrt 2}V_{cb}V_{ud}
a_2 f_D \langle p\bar p|\bar d\centernot p (1-\gamma_5)b|\bar B^0\rangle\,,
\end{eqnarray}
with $G_F$ the Fermi constant, $V_{ij}$ the CKM matrix elements, and
$\centernot p=p^\mu \gamma_\mu$, 
where the decay constants $f_\pi$ and $f_D$ along with the momentum transfer $p^\mu$
come from the matrix elements of 
$\langle \pi|\bar u \gamma^\mu (1-\gamma_5) d|0\rangle=if_\pi p^\mu$ and 
$\langle D|\bar c \gamma^\mu (1-\gamma_5) u|0\rangle=if_D p^\mu$, respectively,
and
$a_i\equiv c^{eff}_i+c^{eff}_{i\pm1}/N_c$
for $i=$odd (even)
are composed of the effective Wilson coefficients $c_i^{eff}$ 
defined in Refs.~\cite{Geng:2006wz,Geng:2006jt,Chen:2008sw,Hsiao:2013dta}.
The matrix elements for $B\to {\bf B\bar B'}$ transition in Eq.~(\ref{amp1}) 
can be parameterized as~\cite{Geng:2006wz}
\begin{eqnarray}\label{transitionF}
&&\langle {\bf B}{\bf\bar B'}|\bar q'\gamma_\mu b|B\rangle=
i\bar u[  g_1\gamma_{\mu}+g_2i\sigma_{\mu\nu}p^\nu +g_3 p_{\mu} 
+g_4q_\mu +g_5(p_{\bf\bar B'}-p_{\bf B})_\mu]\gamma_5v\,,\nonumber\\
&&\langle {\bf B}{\bf\bar B'}|\bar q'\gamma_\mu\gamma_5 b|B\rangle=
i\bar u[ f_1\gamma_{\mu}+f_2i\sigma_{\mu\nu}p^\nu +f_3 p_{\mu} 
+f_4q_\mu +f_5(p_{\bf\bar B'}-p_{\bf B})_\mu]v\,,
\end{eqnarray}
where $g_i(f_i)$ $(i=1,2,3,4,5)$ are
the $B\to{\bf B\bar B'}$ transition form factors, of which
the momentum dependences can be written as~\cite{HouSoni,Chua:2002wn,Geng:2006wz,Geng:2006jt}
\begin{eqnarray}\label{transitionF2}
f_i(t)=\frac{D_{f_i}}{t^3}\;, \qquad g_i(t)=\frac{D_{g_i}}{t^3}\;.
\end{eqnarray}
In terms of the $SU(3)$ flavor and $SU(2)$ spin symmetries,
$D_{f_i}$ and $D_{g_i}$ for different $B\to{\bf B\bar B'}$ transitions can be related, 
given by~\cite{Chen:2008sw}
\begin{eqnarray}\label{C&D}
&&D_{g_1(f_1)}=\frac{5}{3}D_{||}\mp\frac{1}{3}D_{\overline{||}}\,,\;
D_{g_j(f_j)}=\pm\frac{4}{3}D_{||}^j\,,\nonumber\\
&&D_{g_1(f_1)}=\frac{1}{3}D_{||}\mp\frac{2}{3}D_{\overline{||}}\,,\;
D_{g_j(f_j)}=\mp\frac{1}{3}D_{||}^j\,,
\end{eqnarray}
for $\langle p\bar p|\bar u \gamma_\mu(\gamma_5) b|B^-\rangle$ and 
$\langle p\bar p|\bar d \gamma_\mu (\gamma_5)b|\bar B^0\rangle$, respectively,
with the constants $D_{||(\overline{||})}$ 
and $D_{||}^j$ ($j=2,3,4,5$) to be determined by the global fit with
all available data of $B^-\to p\bar p e^-\bar \nu_e$ and $B\to p\bar p M_{(c)}$
($M=\pi,\,K$ and $K^*$ and $M_c=D^{(*)}$).

Subsequently, in terms of the amplitudes in Eq.~(\ref{amp1}), 
we derive $|V_{ub}|^2/|V_{cb}|^2$ as
\begin{eqnarray}\label{R_Vub}
|V_{ub}|^2/|V_{cb}|^2=({\cal B}_\pi/{\cal B}_D){\cal R}_{ff}\,,
\end{eqnarray}
where ${\cal B}_\pi/{\cal B}_D\equiv {\cal B}(B^-\to p\bar p \pi^-)/{\cal B}(\bar B^0\to p\bar p D^0)$,
 and ${\cal R}_{ff}$ 
 is given by
\begin{eqnarray}\label{Rf}
{\cal R}_{ff}=\frac
{\int \int  
|a_2\,f_D\langle p\bar p|\bar d\centernot p (1-\gamma_5)b|\bar B^0\rangle|^2\,dm^2_{p\bar p} dm^2_{\bar p D}}
{\int \int
|a_1\,f_\pi\langle p\bar p|\bar u\centernot p (1-\gamma_5)b|B^-\rangle|^2\,dm^2_{p\bar p} dm^2_{\bar p \pi}}\,,
\end{eqnarray}
with $m^2_{ij}=(p_i+p_j)^2$, 
in which the allowed ranges over the phase space 
can be referred in the PDG~\cite{pdg}.

\subsection{The $|V_{ub}|$ extraction}
Since the relation in Eq.~(\ref{R_Vub}) can be used to extract $|V_{ub}|$,
we adopt the data from PDG as the experimental inputs, given by~\cite{pdg}
\begin{eqnarray}
&&(f_D,\,f_\pi)=(204.6\pm 5.0,130.4\pm 0.2)\;\text{MeV}\,,\nonumber\\
&&{\cal B}(B^-\to p\bar p \pi^-)=(1.60\pm 0.18)\times 10^{-6}\,,\;\nonumber\\
&&{\cal B}(\bar B^0\to p\bar p D^0)=(1.04\pm 0.07)\times 10^{-4}\,,
\end{eqnarray}
which result in ${\cal B}_\pi/{\cal B}_D=(1.54\pm 0.17)\times 10^{-2}$.

For the theoretical inputs, 
we use the new extraction for the $B\to p\bar p$ transition form factors,
which includes the new observation of ${\cal B}(B^-\to p\bar p e^-\bar \nu_e)$~\cite{Tien:2013nga},
such that the overestimation in Ref.~\cite{Geng:2011tr} can be fixed.
The results from the global 
fitting
with all available data of $B^-\to p\bar p e^-\bar \nu_e$ and $B\to p\bar p M_{(c)}$
($M=\pi,\,K$ and $K^*$ and $M_c=D^{(*)}$) 
are given by~\cite{Chen:2008sw,Geng:2011tr,Hsiao:2016amt}
\begin{eqnarray}\label{para}
&&(D_{||},D_{\overline{||}})=(37.1\pm 68.9,-356.5\pm 22.2)\;{\rm GeV}^{5}\,,\nonumber\\ 
&&(D_{||}^2,D_{||}^3,D_{||}^4,D_{||}^5)=
(16.6\pm 30.7,-274.7\pm 171.9, 4.0\pm 29.5,137.8\pm 37.4)\;{\rm GeV}^{4}\,.
\end{eqnarray}
The parameter $a_1$ for the charmless $B^-\to p\bar p \pi^-$ decay
is given by $a_1=c_1^{eff}+c_2^{eff}/N_c$ with $N_c$ the color number,
where the effective Wilson coefficients $c_{1,2}^{eff}$ have been 
adopted to be 
$(c^{eff}_1,\,c^{eff}_2)=(1.168,\,-0.365)$~\cite{Geng:2006wz,Geng:2006jt,Chen:2008sw,Hsiao:2013dta}.
In the generalized version of the factorization, one is able to float $N_c$
from 2 to $\infty$ to estimate the non-factorizable effects, resulting in 
$a_1=1.05\pm 0.12$. Since the parameter $a_2$ for $\bar B^0\to p\bar p D^0$
is sensitive to the non-factorizable effects, the fitting with the all available data 
gives $a_2=0.42\pm 0.04$~\cite{Chen:2008sw,Hsiao:2016amt}. 
We then estimate ${\cal R}_{ff}$ in Eq.~(\ref{Rf}) to be 
\begin{eqnarray}
{\cal R}_{ff}=0.50^{+0.13}_{-0.09}\,,
\end{eqnarray}
which leads to $|V_{ub}|/|V_{cb}|=0.088^{+0.022}_{-0.016}\pm 0.010$
with the errors from ${\cal R}_{ff}$ and ${\cal B}_\pi/{\cal B}_D$, respectively.
Since $|V_{cb}|$ has been well measured, with $|V_{cb}|=(39.5\pm 0.8)\times 10^{-3}$~\cite{pdg},
we obtain
\begin{eqnarray}\label{Result}
|V_{ub}|=
(3.48^{+0.87}_{-0.63}\pm 0.40\pm 0.07)\times 10^{-3}\,,
\end{eqnarray}
with the third error for $|V_{cb}|$. 

\section{Discussions and Conclusions}
Our result in Eq.~(\ref{Result})
is close to the exclusive $\bar B\to \pi \ell\bar \nu_\ell$ and 
$\Lambda_b\to p\mu^-\bar \nu_\mu$ cases; particularly, 
nearly the same as $|V_{ub}|\simeq |A\lambda^3(\rho-i\eta)|\simeq 3.56\times 10^{-3}$
in the Wolfenstein parameterization~\cite{pdg}. Nonetheless, 
the complete estimation of the theoretical uncertainties from the $B\to p\bar p$ transitions
gives the biggest error of 
$0.87\times 10^{-3}$,
such that our result also overlaps
the inclusive value in Eq.~(\ref{Va}).
While the tension between the exclusive and inclusive extractions 
in Eqs.~(\ref{Va}) and (\ref{Vb})
is suspected to be due to the underestimated theoretical uncertainties~\cite{Crivellin:2014zpa},
in our case the range of 
$|V_{ub}|=(2.73-4.43)\times 10^{-3}$
seems to reconcile the difference.

In sum, by relating $B^-\to p\bar p \pi^-$ and $\bar B^0\to p\bar p D^0$ decays,
we have obtained $|V_{ub}|^2/|V_{cb}|^2=({\cal B}_\pi/{\cal B}_D){\cal R}_{ff}$
for the extraction of $|V_{ub}|$, where $|V_{cb}|$ and 
${\cal B}_\pi/{\cal B}_D\equiv{\cal B}(B^-\to p\bar p \pi^-)/{\cal B}(\bar B^0\to p\bar p D^0)$
are given by data, while ${\cal R}_{ff}$ is the parameter related to 
 the $B\to p\bar p$ transition matrix elements.
With ${\cal R}_{ff}=0.50^{+0.13}_{-0.09}$ estimated from the global fitting of
all available data of $B^-\to p\bar p e^-\bar \nu_e$ and $B\to p\bar p M_{(c)}$
($M=\pi,\,K$ and $K^*$ and $M_c=D^{(*)}$),
we have found that $|V_{ub}|=(3.48^{+0.87}_{-0.63}\pm 0.40\pm 0.07)\times 10^{-3}$,
where the errors correspond 
to the uncertainties from ${\cal R}_{ff}$, ${\cal B}_\pi/{\cal B}_D$ and $|V_{cb}|$, respectively.
Being independent of the previous results,
the extraction of $|V_{ub}|$ in the baryonic $B$ decays is clearly very useful for the complete 
determination of the CKM matrix elements as well as the exploration of new physics.

\section*{ACKNOWLEDGMENTS}
We are grateful  for the useful communications from Professor Robert Kowalewski.
The work was supported in part by National Center for Theoretical Sciences, National Science
Council (NSC-101-2112-M-007-006-MY3), MoST (MoST-104-2112-M-007-003-MY3) and National Tsing Hua
University (104N2724E1).

\end{document}